\documentclass[10pt,a4paper,twocolumn]{article}
\usepackage{titlesec}
\usepackage{color}
\usepackage[utf8]{inputenc} 
\usepackage[affil-it]{authblk}

\usepackage{lipsum}
\newcommand\blfootnote[1]{%
  \begingroup
  \renewcommand\thefootnote{}\footnote{#1}%
  \addtocounter{footnote}{-1}%
  \endgroup
  }
\usepackage{siunitx}
\usepackage{float}
\usepackage{tikz}
\usepackage{xcolor}

\usepackage{braket}
\usepackage{nicefrac}
\usepackage[english]{babel} 
\usepackage[top=2.8cm,bottom=2.8cm,left=2.cm,right=2.cm]{geometry}  
\usepackage{amsmath,amsfonts,amssymb}  

\usepackage{graphicx,subfig}  
\graphicspath{{./Figures/}{./PSTricks/}}  
\usepackage{multirow}
\usepackage{appendix}
\usepackage{wrapfig}
\usepackage{setspace}
\usepackage{cite}

\usepackage{csquotes}

\usepackage{array,booktabs}

\usepackage{enumitem}
\usepackage{caption3} 
\DeclareCaptionOption{parskip}[]{}
\usepackage[font=small,labelfont=bf]{caption}
\usepackage[final]{pdfpages}
\usepackage[colorinlistoftodos]{todonotes}
\usepackage{tikz}
\usepackage{sidecap}
\sidecaptionvpos{figure}{t}
\usepackage{abstract}

\usepackage[pdftex]{hyperref} 

\title{\textbf{High-Dimensional Quantum Key Distribution based on Multicore Fiber using
Silicon Photonic Integrated Circuits}} 
\author{Yunhong Ding$^1$*, Davide Bacco$^1$$^\dagger$, Kjeld Dalgaard$^1$, Xinlun Cai$^2$, Xiaoqi Zhou$^3$, Karsten Rottwitt$^1$, and Leif Katsuo Oxenløwe$^1$}
\affil{$^1$ \small Department of Photonics Engineering, Technical University of Denmark, 2800 Kgs.~Lyngby, Denmark.}
\affil{$^2$ School of Electronics and Information Technology, State Key Laboratory of Optoelectronic Materials and Technologies, Sun Yat-sen University, Guangzhou, China }
\affil{$^3$ School of Physics and Engineering, State Key Laboratory of Optoelectronic Materials and Technologies, Sun Yat-sen University, Guangzhou, China }
\date{\vspace{-1em} \small   Dated: \today }

\begin{document}


\pagestyle{plain}
\setcounter{page}{1}
\twocolumn[ 
\begin{@twocolumnfalse}
\maketitle
     \vspace{-0.8cm}
  \begin{abstract}
      \normalsize
         \vspace*{-1.0em}
\noindent Quantum Key Distribution (QKD) provides an efficient means to exchange information in an unconditionally secure way. 
Historically, QKD protocols have been based on binary signal formats, such as two polarisation states, and the transmitted information efficiency of the quantum key is intrinsically limited to 1 bit/photon. Here we propose and experimentally demonstrate, for the first time, a high-dimensional QKD protocol based on space division multiplexing in multicore fiber using silicon photonic integrated lightwave circuits. We successfully realized three mutually unbiased bases in a four-dimensional Hilbert space, and achieved low and stable quantum bit error rate well below both coherent attack and individual attack limits. Compared to previous demonstrations, the use of a multicore fiber in our protocol provides a much more efficient way to create high-dimensional quantum states, and enables breaking the information efficiency limit of traditional QKD protocols. In addition, the silicon photonic circuits used in our work integrate variable optical attenuators, highly efficient multicore fiber couplers, and Mach–Zehnder interferometers, enabling manipulating high-dimensional quantum states in a compact and stable means. Our demonstration pave the way to utilize state-of-the-art multicore fibers for long distance high-dimensional QKD, and boost silicon photonics for high information efficiency quantum communications.
\end{abstract}
  \end{@twocolumnfalse}
 ]

\begin{figure*}
   \centering
    {\includegraphics[width=17cm]{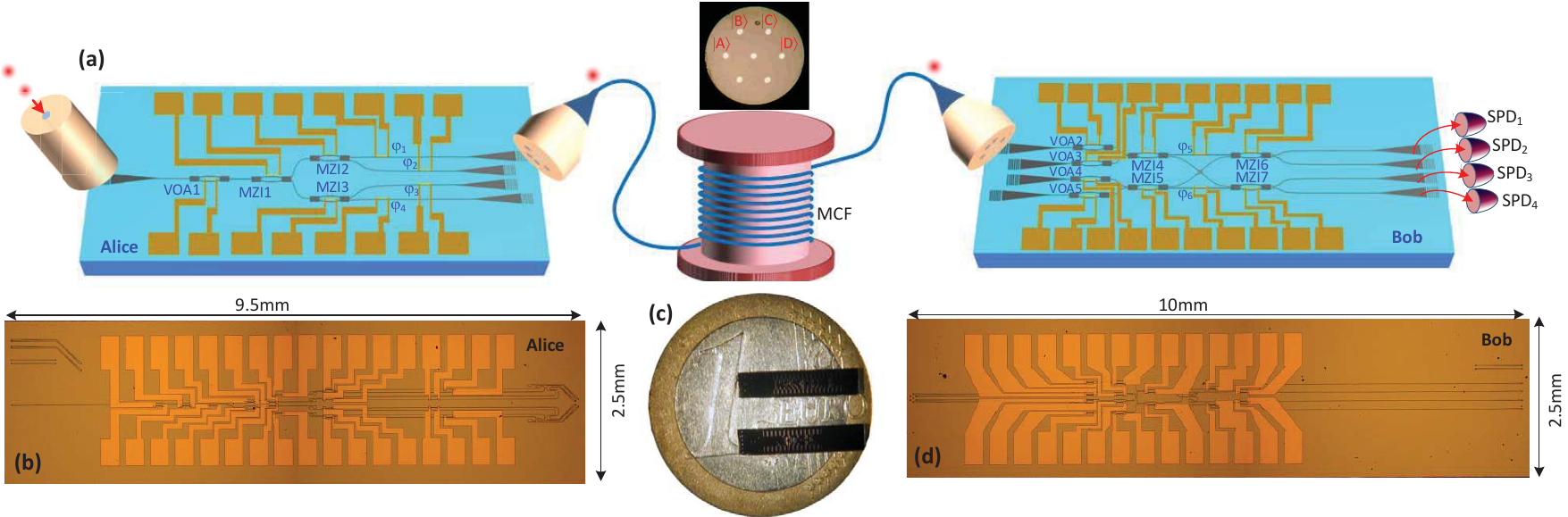} }
    \caption{(a) Schematic of the HD-QKD based on MCF using silicon PIC for Alice and Bob. The inset shows the cross-section of the multi-core fiber, where four cores are used. (b) and (d) shows the fabricated silicon PIC for Alice and Bob, respectively. (c) presents the picture of the fabricated chips with a 1 euro coin, indicating the compact size of the silicon PICs.}
\label{fig:Fig1}
\end{figure*}

\subsection*{Introduction}
Quantum\blfootnote{* yudin@fotonik.dtu.dk, $^\dagger$dabac@fotonik.dtu.dk} key distribution (QKD) is an attractive quantum technology that provides a means to securely share secret keys between two clients (Alice and Bob)~\cite{Lo1999,Hwang2003,Bacco2012,Lo2014}. Traditional QKD is based on binary signal formats, such as the BB84 protocol where the quantum information is encoded in the polarization domain~\cite{BBPr}. Four polarization states create a set of two mutually unbiased basis (MUBs) in a two-dimensional Hilbert space which are used for establishing quantum keys between two parties. 
In these binary QKD systems the information efficiency is limited to $1$ bit/photon. Recently, tremendous efforts have been put into developing on novel protocols to increase the information efficiency, because protocols based on qudit encoding (unit of information in a $d$ dimension space) exhibit a higher resilience to noise, allowing for lower signal-to-noise ratio (SNR) of the received signal, which in turn may be translated into longer transmissin distances~\cite{Mirh,Cotler,Bacco2016,Mafu2013,cerf2002}. High-dimensional QKD (HD-QKD) is an efficient means to achieve high information efficiency of QKD systems ~\cite{Etcheverry2013,Mower2013,Bunandar2015,Zhong2015,walborn2006}. 
One interesting way to achieve HD-QKD is to use space division multiplexing (SDM) technology, where the spatial dimension is used to carry the quantum states. 
In this context, optical angular momentum (OAM) modes has been proposed as a HD-QKD protocol, and demonstrated over a free-space link using discrete components. QKD systems will eventually be merged with optical fiber transmission, so as to link the classical internet with a quantum internet~\cite{Paterson2005,Groblacher2006,DAmbrosio2012,Vallone2014}. 
However, the transmission of OAM modes over long distance fiber links is very challenging due to inter-modal crosstalk, which is a critical issue for HD-QKD protocol. 
An alternative approach is to use separate cores in multicore fibers (MCFs). MCFs have been widely studied for classical communication systems, resulting in ultra-high communication capacity, owing to the large potential multiplicity of cores and the low crosstalk between cores ~\cite{Mizuno2016}.  
In addition, it seems clear that in order to develop the next generation of QKD systems, integrated quantum transmitters and receivers must be considered. Integrated circuits provides excellent optical phase stability, making them particularly suitable for manipulating quantum states in compact chips with low energy consumption. Binary QKD systems have been demonstrated using integrated circuits ~\cite{Sibson2015,Ma2016}, showing comparable performance compared to the bulky systems. 
Within integrated optics, silicon photonics technology has been a powerful means to combine the assets of integrated photonics with complementary metal-oxide semiconductor (CMOS) technologies. In this paper, we demonstrate the first HD-QKD protocol based on multicore fibers (MCFs) using silicon photonic integrated circuits (PICs). We have proved that manipulation of the high dimensional quantum states in MCFs is feasible. We successfully prepared three mutually unbiased bases with four dimensional quantum states and sent these through a 7-core fiber. Good extinction ratio is obtained in tomography measurements. Stable and low quantum bit error rate (QBER), below threshold limits, is achieved for more than 10 minutes.

\section*{Results}
\subsection*{Protocol definition}
Most of the QKD protocols are based on the concept of MUBs. For instance, $\lbrace \mathtt{B_0} = \ket{\Phi_{0,i} \, i=0,1,\dots, N-1} \rbrace$ and $\lbrace \mathtt{B_1} = \ket{\Phi_{1,j} \, j=0,1,\dots, N-1} \rbrace$ are orthogonal bases of a N-dimensional Hilbert space $H_\mathbf{N}$. These bases are defined as mutually unbiased if and only if 
\begin{equation}
\lvert \langle \Phi_{0,i} \vert \Phi_{1,j} \rangle \rvert^2 =\frac{1}{N}
\end{equation}
In other words in a set of MUBs $\lbrace \mathtt{B_0},\mathtt{B_1}, \mathtt{B_2}, \dotso, \mathtt{B_k}, \mathtt{B_n} \rbrace$ if we measure a state in $\mathtt{B_n}$ basis, and this state was prepared in  $\mathtt{B_k}$ basis (with $k \neq j$), all the outputs are equally probable. It has been proved that the maximum number of the MUBs in a Hilbert space of dimension $N$ is $N+1$, where $N$ is a integer power of a prime number{\cite{boykin2002}.
We restrict our analysis to the case of three MUBs reported in equation~\ref{eq:mubs} for an Hilbert space of four dimension ($N=4$). The main difference is related to the tolerable threshold of the quantum bit error rate (QBER) and the maximum achievable value of secret key rate. A more detailed discussion will follow in next paragraph.
The set of the three MUBs, used in the proposed HD-QKD system, can be defined as a linear combination of:
\begin{align*}
\ket{A}&= \frac{1}{\sqrt{2}} \: ( 1 \, 0 \, 0 \, 0)\\
\ket{B}&= \frac{1}{\sqrt{2}} \: ( 0 \, 1 \, 0 \, 0)\\
\ket{C}&= \frac{1}{\sqrt{2}} \: ( 0 \, 0 \, 1 \, 0)\\
\ket{D}&= \frac{1}{\sqrt{2}} \: ( 0 \, 0 \, 0 \, 1)
\end{align*} 
\begin{small}
\begin{equation}
\mathtt{M_0}=
\begin{pmatrix}
  \ket {A} + \ket {B}\\
  \ket {A} - \ket {B}\\
  \ket {C} + \ket {D}\\
  \ket {C} - \ket {D}
\end{pmatrix}
\mathtt{M_1}=
\begin{pmatrix}
  \ket {A} + \ket {C}\\
  \ket {A} -\ket {C}\\
  \ket {B} + \ket {D}\\
  \ket {B} - \ket {D}
\end{pmatrix}
\mathtt{M_2}=
\begin{pmatrix}
  \ket {A} + \ket {D}\\
  \ket {A} - \ket {D}\\
  \ket {B} + \ket {C}\\
  \ket {B} - \ket {C}
\end{pmatrix}
\label{eq:mubs}
\end{equation}
\end{small}
\noindent In this way we can easily define the set ${\mathtt{M_0}, \mathtt{M_1}, \mathtt{M_2}}$, that respects the mutually unbiased assumption, $ \lvert \langle \mathtt{M_0}\vert \mathtt{M_1} \rangle \rvert^2 = \: \lvert \langle \mathtt{M_0}\vert \mathtt{M_2} \rangle \rvert^2 = \: \lvert \langle \mathtt{M_1}\vert \mathtt{M_2} \rangle \rvert^2 = 1/4$.
Physically, \textbar A$\rangle$, \textbar B$\rangle$, \textbar C$\rangle$, and \textbar D$\rangle$ represent the quantum states related to the four individual cores, as shown in the inset of Figure ~\ref{fig:Fig1} (a). By tuning the Mach-Zehnder interferometers (MZIs), situated in the transmitter chip, Alice creates a quantum superposition between cores. In this way she prepares one of the states in the three MUBs. A random number generator is used for basis and states choice.  Before the quantum measurement, Bob randomly choose one of the three MUBs tuning the corresponding MZI.
In such a way, Bob creates interference between different cores and he correctly measures the quantum states sent by Alice. As in the BB84 protocol, after the measurements, a distillation process is required. In this procedure Alice and Bob discard all the data related to a different basis choice. At this point Alice and Bob share an identical quantum key, useful for encryption and decryption of the plain text.

\subsection*{Secret key rate}
One of the most important parameters in a communication system is the achievable rate. In particular, in a QKD system, the major criterion is represented by the secret key rate: number of bit/s or bit/pulse that Alice and Bob can create as useful key. A general formula for the standard protocols can be written as:
\begin{equation}
R \: \geq \: I_{AB}- min(I_{AE},I_{BE})
\label{eq:rate}
\end{equation}
where $I_{AB}$ represents the classical mutual information between Alice and Bob ($I_{XY} = H(X)-H(X \vert Y)$), and the marginal entropy is defined as $H(X) = \sum_{x \in X} {p(x) \log p(x)}$. The right term of equation~\eqref{eq:rate} $\min$($I_{AE}$ and $I_{BE}$), are related to the quantum mutual information between Alice and Eve or Bob and Eve. In the following analysis we take into account only the mutual information between Alice and Eve, but a more complete analysis can be done. Moreover, we also make the assumption of trusted-device scenario, in which Eve cannot modify the efficiency of Bob’s detectors. Let us now define the secret key rate formula in a four-dimensional QKD system. The mutual information between Alice and Bob is: 
\begin{equation}
I_{AB}= log_2(N) + F \, log_2 (F) + (1-F) log_2 (\frac{1-F}{N-1})
\label{eq:iab}
\end{equation}
where $N$ is the dimension of the Hilbert space and $F=1-D$ represents the fidelity of Bob. $D$ is defined as the disturbance in the communication link.
In order to extract a bound on the final secret key rate, we should define the best strategy for Eve.  In the following analysis we consider two different kinds of Eve's strategy. Individual attacks (IAs), where Eve monitors the ququarts independently from each other, and General attacks (GAs). GAs instead are less conservative on Eve's strategy, in fact she can monitor more quantum states jointly. Intuitively GAs are less stringent than IAs.

\noindent Let's start our analysis under the assumption of IAs where Eve uses a universal quantum cloning machine for qudits. Focusing on the $2$ MUBs we retrieve the mutual information between Alice and Eve depending on the number of states $N$.
In the case of ququarts encoding we define four different quantum states as $\ket{0}, \ket{1}, \ket{2}, \ket{3}$.
The most general symmetric eavesdropping strategy for ququarts can be written as:
\begin{align*}
\ket{0} \ket{E} \: & \xrightarrow{\mathcal{U}} \: \sqrt{1-D} \ket{0} \ket{E_{00}} + \sqrt{\frac{D}{N}} \ket{1} \ket{E_{01}}\\ 
& \hspace{1cm}+ \sqrt{\frac{D}{N}} \ket{2} \ket{E_{02}} + \sqrt{\frac{D}{N}} \ket{3} \ket{E_{03}}\\ 
\ket{1} \ket{E} \: & \xrightarrow{\mathcal{U}} \: \sqrt{\frac{D}{N}} \ket{0} \ket{E_{10}} + \sqrt{1-D} \ket{1} \ket{E_{11}}\\ 
& \hspace{1cm} + \sqrt{\frac{D}{N}} \ket{2} \ket{E_{12}} + \sqrt{\frac{D}{N}} \ket{3} \ket{E_{13}}\\
\ket{2} \ket{E} \: & \xrightarrow{\mathcal{U}} \:  \sqrt{\frac{D}{N}} \ket{0} \ket{E_{20}} 
 + \sqrt{\frac{D}{N}} \ket{1} \ket{E_{21}}\\
& \hspace{1cm} + \sqrt{1-D} \ket{2} \ket{E_{22}} + \sqrt{\frac{D}{N}} \ket{3} \ket{E_{23}} \\
\ket{3} \ket{E} \: & \xrightarrow{\mathcal{U}} \:  \sqrt{\frac{D}{N}} \ket{0} \ket{E_{30}} 
 + \sqrt{\frac{D}{N}} \ket{1} \ket{E_{31}}\\
& \hspace{1cm} + \sqrt{1-D} \ket{2} \ket{E_{32}} + \sqrt{\frac{D}{N}} \ket{3} \ket{E_{33}} \end{align*}
Under the assumption of symmetric attacks (Eve treats all the input states with the same strategy) and following the approach defined in~\cite{cerf2002} the fidelity is: $F=Tr (\ket{\psi_i} \bra{\psi_i} \rho_B^{out}) $, where $\rho_B^{out}$ is the reduced density operator of the state send to Bob. The maximum Eve's mutual information is: 
\begin{equation}
I_{AE}= log_2(N) + F_E \, log_2 (F_E) + (1-F_E) log_2 (\frac{1-F_E}{N-1})
\label{eq:iae}
\end{equation}
where the corresponding optimal fidelity for Eve is:
\begin{multline}
F_E = \frac{F}{N} + \frac{(N-1)(1-F)}{N} \\
+ \frac{2}{N} \sqrt{(N-1) F (1-F)}
\end{multline}
In other words, equation~\eqref{eq:iae} represents the ability of Eve to distinguish between $N$ non-orthogonal states . In the case of IAs $I_{AE}$ can be easily formulated using $F_E$ instead of $F$ in equation~\eqref{eq:iab}.
Using equation~\eqref{eq:rate} we can define an upper bound on the disturbance $D$ introduce by Eve. We highlight that this limit increases with the Hilbert space dimension (see supplementary information Table S.1). As a result, the higher $N$ used in the system, the higher QBER is allowed in the key generation (see Figure~\ref{fig:simulation} (a) ).
Following the same approach it is possible to defined a bound on Eve's information related to the case of $N+1$ MUBs~\cite{cerf2002}.
Let's now consider a different strategy for Eve. As introduced above in case of CAs Eve collects a finite number of ququart pulses, $n<\infty$. In this case under the assumption made in~\cite{cerf2002} and~\cite{hall1997}, it is possible to define an upper bound on Eve's information. The condition
$
I_{AB} \geq \log_2 (N)
$ is a sufficient case for guaranteeing the security under CAs, if the key dimension $l$  is much greater than $n$~\cite{hall1997}. All the bounds presented are true under the assumption of infinity key length, but different approach can be used for defining security under the case of finite key scenario. 

\noindent In the last part of this paragraph we introduce the concept of HD-Decoy state QKD. The idea of Decoy state QKD was introduced by {\it Lo et al}. in 2005\cite{lo2005}. Most of the QKD systems use different degrees of freedom like polarization, phase and time carried by a train of weak coherent pulses (WCPs). This idea, although unconditionally secure from a theoretical point of view, will encounter some limitation on practical realization. In fact, WCPs are obtained by attenuated laser with a mean photon number per pulse, $\mu$, lower than one. However, the Poissonian distribution of the laser not guarantee a non zero probability of multi-photon states. In that case, Eve can easily stop all the single photon pulses and split the multi-photon pulses. In such a way Eve can measure the same information of Bob and after information reconciliation Alice, Eve and Bob will share the same key.
In order to overcome this problem a simple strategy can be adopted. During the transmission process, Alice changes randomly the mean photon number per pulse. A modulable intensity laser or a real time variable optical attenuator can be used to create different levels of mean photon numbers.
As reported in~\cite{lo2005} two different values, and a vacuum measurement, are sufficient to guarantee a good compromise between rate and security.


\begin{figure*}
   \centering
    \includegraphics[width=16cm]{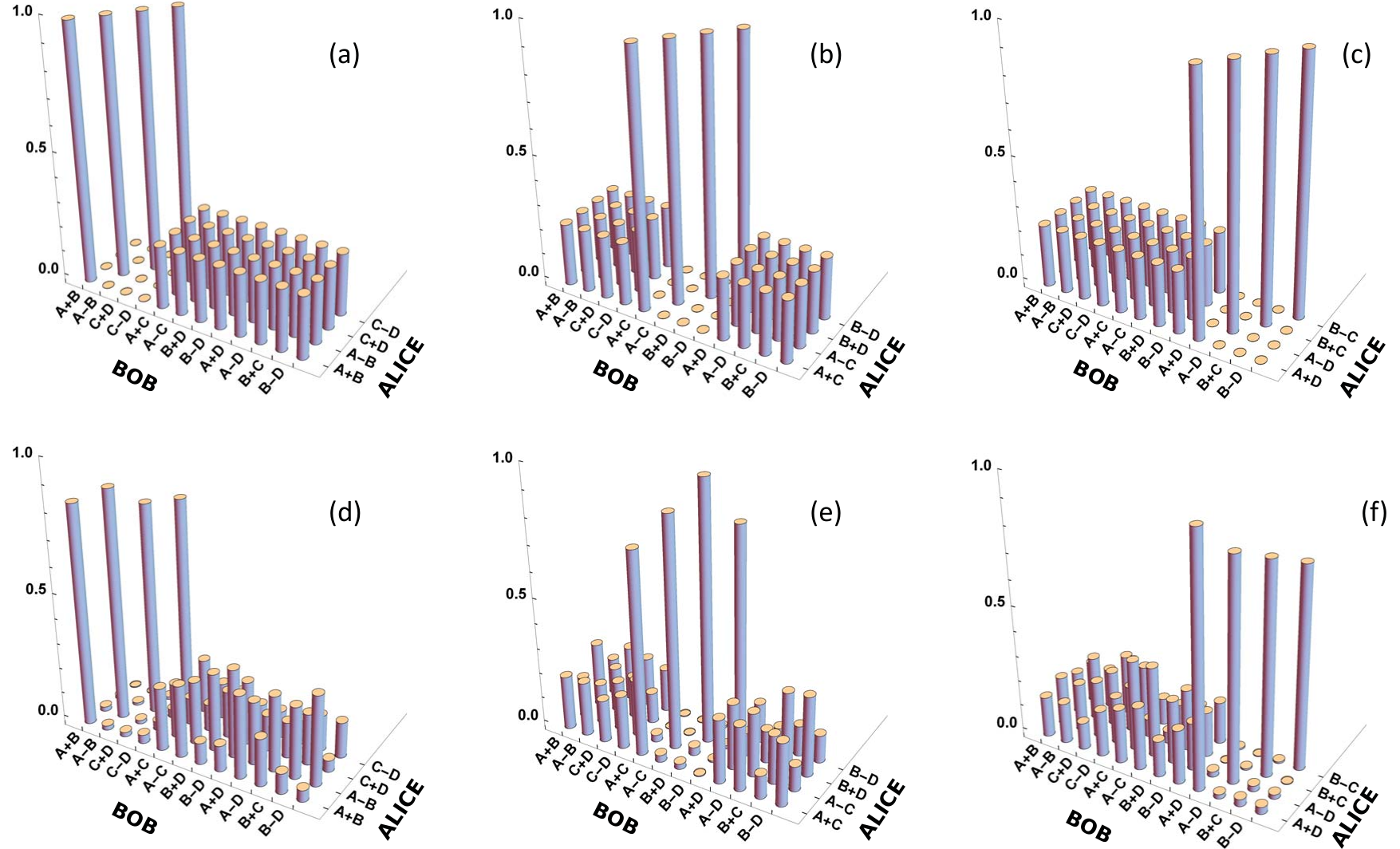}
    \caption{State tomography: figures (a), (b), (c) represent the theoretical matrices of the different MUBs; figures (d), (e), (f) show the experimental measurements of the same basis. The state tomography were obtained with $30$ seconds of integration time using a CW $1550$ nm laser modulated with a $10$ ns pulse at $10$ kHz of repetition rate and a mean photon per pulse of $0.2$.}
    \label{fig:tomography}
\end{figure*}

\subsection*{Experimental results}

The experimental scheme of the proposed HD-QKD based on MCF using silicon PICs is shown in Fig. 1(a) (the detailed experimental setup is presented in supplementary information, section 1). WCPs (10 ns wide, see also supplementary information) are injected into the transmitter chip (Alice, Fig.~\ref{fig:Fig1} (b)). The silicon PIC for Alice is used to prepare the high-dimensional quantum states. A variable optical attenuator (VOA1) is used to thermally tune the pulse power to prepare the decoy states. Cascaded MZIs associated with the four thermally tunable phase  shifters are used to prepare the quantum states in the three MUB sets with mean number of photons per pulse, $\mu$, lower than $1$. As an example, MZI 1 can be set so that light goes to the upper arm. At the same time, MZI 2 can be set as a 3 dB coupler. In this case, $M_0$ basis is prepared, and states $\ket{A}+\ket{B}$
and $\ket{A} - \ket{B}$ can be prepared by controlling the phase difference between the two output arms of MZI 2 to $0$ or $\pi$, which is obtained through thermally controlling $\phi 1$ and  $\phi 2$. Similarly, the other quantum states can be prepared by configuring the corresponding MZIs and phase shifters. The total insertion loss of Alice chip is 15 dB, which includes the coupling loss with the input single mode fiber (SMF) and the MCF fan-in/fan-out (FI/FO), and all the losses on the chip. Alice randomly chooses one of the four states in one of the three bases to transmit to Bob. 
By tuning the VOA1, Alice introduces a 10 second of decoy state sequence every 2 minutes. In this way we can prepare in real time the signal and decoy quantum states.
The quantum states are coupled to four cores of a multi-core fiber through an apodized grating coupler based MCF FI/FO~\cite{Ding2013,Ding2015}. After the MCF, the quantum states are coupled into Bob’s chip (Fig.1(d)) through the MCF FI/FO coupler, and randomly measured in one of the three bases by configuring the corresponding phase shifters and MZIs. For instance, by configuring MZI 4 as a 3 dB coupler and setting MZI 6 and MZI 7 to directly go through, Bob will be configured to receive the quantum states in base $M_0$. In order to measure correctly the quantum states, it is critical to obtain balanced losses for the four spatial channels, which is enabled by four VOAs introduced in BOB's chip. The total insertion loss of Bob chip is 8 dB. As a proof of concept, 3 meters of MCF is used. In the succeeding distillation process, counts measured in the wrong basis are discarded. 

Fig. 2(a) and 2(b) show the theoretically and the experimental quantum tomography of the three MUB. In the measurements, a Weak coherent pulse with $\mu <0.1$ at 10 kHz was injected to Alice, and Bob acquires data for 30 seconds. The good agreement between the theoretical and experimental results indicate that the three MUBs are well prepared. Fig. 3(a) presents the quantum bit error rate (QBER) as a function of time when Alice randomly sends quantum states with a repetition rate of $5$ kHz (currently limited by the on-chip thermal heaters, see supplementary information, section $2$) and $\mu =0.26$ photon per pulse. The corresponding sifted bits transmitted from Alice as a function of time is presented in Fig. 3(b), clearly indicating an introduction of decoy states with sifted bits of $~11$, corresponding to a $\mu$ of $0.15$. Good and stable QBER performance with an average of $13\%$ is achieved for more than 10 minutes, which is well below both coherent attach limit and individual attach limit.

\begin{figure}
\begin{center}
\includegraphics[width=8cm]{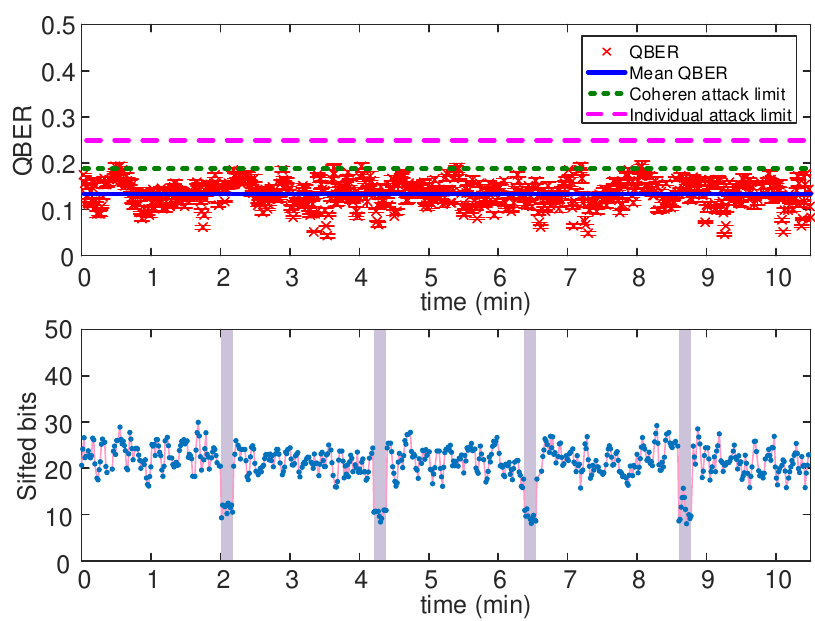}
\end{center}
\caption{\textit{Experimental Results}. (a) QBER  for $10$ minutes of acquired data. The blue line represents the average QBER about $13\%$, which is well below the limits of both coherent attach and individual attacks.  (b) Sifted bits as a function of time. We reported the sifted bits after basis reconciliation for signal and decoy state. The shadow pink area represents the decoys state signal. Controlling the voltage of a VOA1 in Alice chip, it was possible to decrease the mean number photon pulse. Average $\mu$ and $v$ estimation correspond to $0.26$ and $0.15$ photon/pulse.}
\label{fig:ex_results}
\end{figure}

\subsection*{Discussion} 
In the following we discuss results obtained in our HD-QKD experiment and we compared our demonstration with SOTA QKD systems,
The results reported in Figure~\ref{fig:ex_results} (a) and (b) show a stable and good demonstration of the HD-protocol. The average value of $13 \%$ of the QBER, obtained over 10 minutes of measurement, represents a good point for the key extraction. Moreover, thanks to the on-chip VOA1, we implemented a real-time decoy state technique. We highlight this strategy in the shadow pink area of Figure~\ref{fig:ex_results} (b), where a different number of received bits are reported.
The real time experiment was performed using two of three MUBs prepared and showed in Figure~\ref{fig:tomography}. We would like to highlight that the maximum number of MUBs in an Hilbert space of four-dimension is $5$. However, using $5$ MUBs we will compromise the final secret key rate. In fact, although the security increases using more bases, the secret key rate (equation~\eqref{eq:rate}) scales dependently with the number of the bases. As reported in Table 1 of the Supplementary information, for a space of four-dimension, less than $2\%$ of disturbance can be gained in the QBER threshold respect to a leak of a factor $2/5$ on the key rate. The user must choose a trade off between security and rate, depending on the actual conditions of the channel.  


It is well known in optical classical communication, that phase and polarization of coherent light is changed in a long optical fiber transmission. This effect is mainly due to the environment changes during transmission. In a MCF, each core acts independently from the other with very low crosstalk. However the phase and polarization of photons drift independently in each core. In our experiment this effect was mitigated by the short link distance, but in a real QKD system it must be considered. In order to avoid and mitigate this phenomena, two different strategies can be used. Firstly, a two dimensional grating coupler associated with an MZI ~\cite{Caspers2014} can be used to couple with the cores of the MCF, so that the polarization for each core can be tuned independently. Moreover, the polarization dependence can be solved by an on-chip polarization diversity circuit ~\cite{Bogaerts2007,VanLaere2009} for Bob's chip. In addition, Bob is performing interference between cores for receiving quantum states in different basis, thus the time delay between each cores should be matched after long transmission. This problem can be solved by introducing a cascaded delay~\cite{Sibson2015} on Alice chip. An active feedback loop can also be utilized to stabilize phase for each core.

The pulse rate used in this experiment of $10$kHz and $5$kHz represents the main limitation for a long link deployment of this technology. The main limitation is the slow thermal dynamics of the heaters used in our experiment. In fact, despite a very high extinction ratio and good stability, the thermal tunable MZIs on Alice's and Bob's chips have low switching time of 66 $\mu$s and 27 $\mu$s for rise and fall time, as detailed in supplementary information. In order to overcome this problem, two main ideas can be implemented. Different kinds of material, such as graphene, that has ultra-high thermal conductivity enabling sub-microsecond tuning speed can be investigated ~\cite{Gan2015,Yan2016}. Ultra-high speed p-i-n or p-n junction ~\cite{Png2004,Liu2007} can also be utilized for the phase shifters to further increase the repetition rate. The insertion loss of receiver chip is critical for final key rate. Our current device has an insertion loss of 8 dB, which attributes from the coupling loss of the grating couplers. Significant improvements will be achieved by introducing aluminum (Al) mirror below the grating coupler by the flip-bonding method ~\cite{Ding2014,Ding2016}.

Once these technical problems is solved, the idea of space multiplexing HD-QKD can be considered feasible for an end-to-end high-rate transmission over long distance. Moreover, thanks to the advantages of state of the art silicon photonics, laser sources can be integrated on Alice chip in order to create a miniaturized quantum transmitter. Transistors, switches, DACs and other electronic devices can also be integrated on a silicon substrate creating a stable and powerful chip. On the receiver side, fully demonstration of single photon detector on chip was already proved during the last years\cite{tosi2014}. The best solution for HD-QKD is based on single photon arrays. In this way the scalability of the process can be improved and higher capacity protocols can be implemented.

\begin{figure}[!ht]
\subfloat[Information probability]{{\includegraphics[width=9cm]{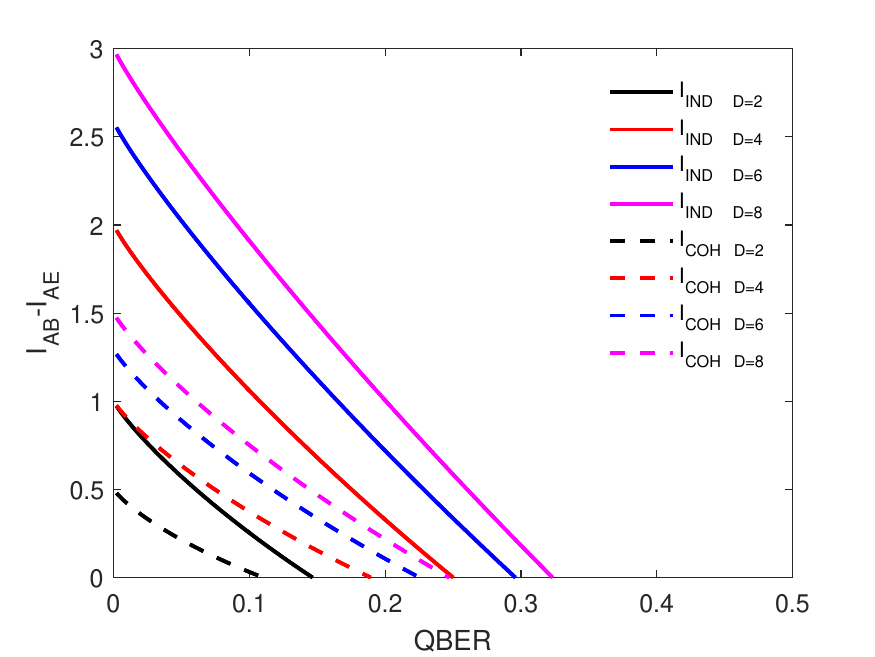}}} \qquad
\subfloat[Secret key rate]{{\includegraphics[width=9cm]{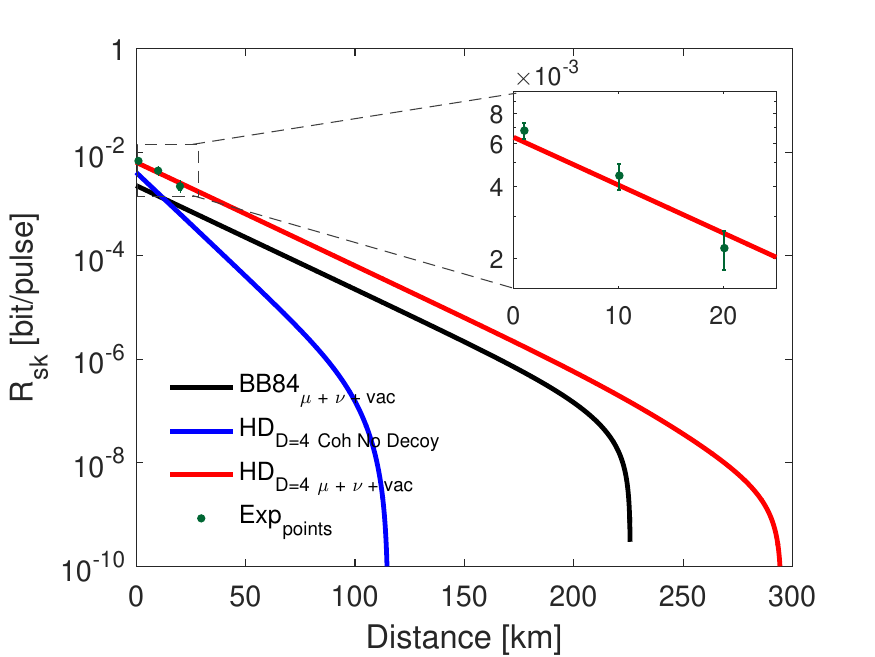}}}\\
\caption{\textit {Simulation results using HD-QKD protocols.} In (a) the mutual information thresholds between Alice and Bob/Eve in case of different Hilbert space dimensions and different Eve's strategy. We considered the case of $2$ MUBs with individual and coherent attacks.  In (b) we reported the simulation of the secret key rate versus the distance. The parameters used are: $p_{dark}= 2 \times 10^{-8}$, $\eta_d=0.1$, $\alpha_{loss} =0.2$ dB/km, $u=0.45$, $v=0.15$ for the decoy rates. Blue solid line is the rate without decoy state in case of WCP HD-QKD. Red and black solid lines are the simulated decoy state key rate for qubit or ququart encoding. We also displayed three experimental points obtained by simulating the corresponding attenuation of the channel. The high-dimension QKD protocols results are more robust against losses and permits a higher distance.}
\label{fig:simulation}
\end{figure}

\subsection*{Methods}
\textbf{Device fabrication}
The silicon PICs was fabricated on the commercial SOI wafer with top silicon thickness of 250 nm and buried oxide layer (BOX) of 1  $\mu$m. A single step of standard SOI processing, including e-beam lithography and inductively coupled plasma (ICP) etching was first used to fabricate the whole silicon PIC simultaneously. A 1500 nm thick layer of SiO2 was then deposited on top of the chip by plasma-enhanced chemical vapor deposition. The chip surface was then polished, and the top SiO2 was thinned down to 1 $\mu$m accordingly. The 1 $\mu$m SiO2 was used as an isolation layer between the silicon waveguide and the Ti heaters fabricated later to avoid potential optical losses. Afterwards, the 100 nm thick titanium heaters are formed by e-beam lithography followed by metal deposition and liftoff process. After that, the thick Au/Ti contact layer was fabricated by UV lithography followed by metal deposition and liftoff process. The chip was then cleaved and wire-bonded to a PCB board, and controlled by FPGA (Field programmable gate array) for system experiment.

\textbf{Electronic design}
The chip-to-chip HD-QKD based on space-division multiplexing is feasible thank to a real time control of the different MZIs presented on the silicon chip. These MZIs, as reported above, are controlled by heaters: conductor material which change his property when a voltage is applied. In order to tune in real-time these MZIs, different electrical signals are required in the transmitter and receiver side. An Altera FPGA board emits $8$ digital parallel outputs every $0.2$ ms, which are converted into analog voltages by $8$ digita-analog converters (DACs). Then, these analog signals are send to the transmitter and the receiver PCB board by flat cables.  

\subsection*{Acknowledgements}
This work is supported by the Danish Council for Independent Research (DFF-1337-00152 and DFF-1335-00771) and the Center of Excellence, SPOC (Silicon Photonics for Optical Communications, ref DNRF123).

\subsection*{Author contributions}
Y. Ding, D. Bacco, X. Cai, and X. Zhou proposed the idea. Y. Ding designed and fabricated the silicon PICs. K. Dalgaard designed the electrical controlling circuits. Y. Ding, D. Bacco, and K. Dalgaard performed the system experiment. D. Bacco carried out the theoretical analysis on the proposed protocol. Y. Ding, D. Bacco, K. Rottwitt, and L. K. Oxenløwe discussed the results. All authors contributed to the writing of the manuscript.

\subsection*{Competing financial interests}
The authors declare no competing financial interests.


\titleformat*{\section}{\bf\normalsize}

\bibliographystyle{abbrv}

\cleardoublepage
\onecolumn
\setcounter{figure}{0}
\renewcommand{\figurename}{Figure S.\!}
\renewcommand{\tablename}{Table S.\!}

\subsection*{Supplementary Information}

\subsection*{Experimental setup}
In Fig.S~\ref{fig:setup} we reported the experimental setup used for the HD-QKD experiment and characterization of the quantum states. We used an CW (continuous-wave) laser ($1540-1560$ nm) curved by an IM (Intensity modulator), which was controlled by FPGA, in order to create a pulsed stream at $5$ kHz repetition rate. The polarization of the stream of pulses was controlled by a polarization controller, and injected in the Alice chip on the TE mode using a fiber grating coupling. The output quantum states from Alice chip was coupled out to the MCF by apodized grating coupler based MCF FI/FO. Both the coupling angle and rotation of the output MCF were optimized in order to get good alignment to the MCF FI/FO. After transmitting over the MCF, the quantum states were coupled to the Bob chip through MCF FI/FO, and detected in one of the prepared bases. Quantum state preparing in Alice and receiving quantum states in Bob were all controlled by the FPGA. The outputs of four cores of the MCF was detected by four SPDs, whose outputs were further connected to the time-tagger. In addition, a low repetition rate synchronization pulse was generated from the FPGA, and injected into the time-tagger. The detected counts were finally analyzed in a computer.

\begin{figure}[!b]
\includegraphics[width=1\textwidth]{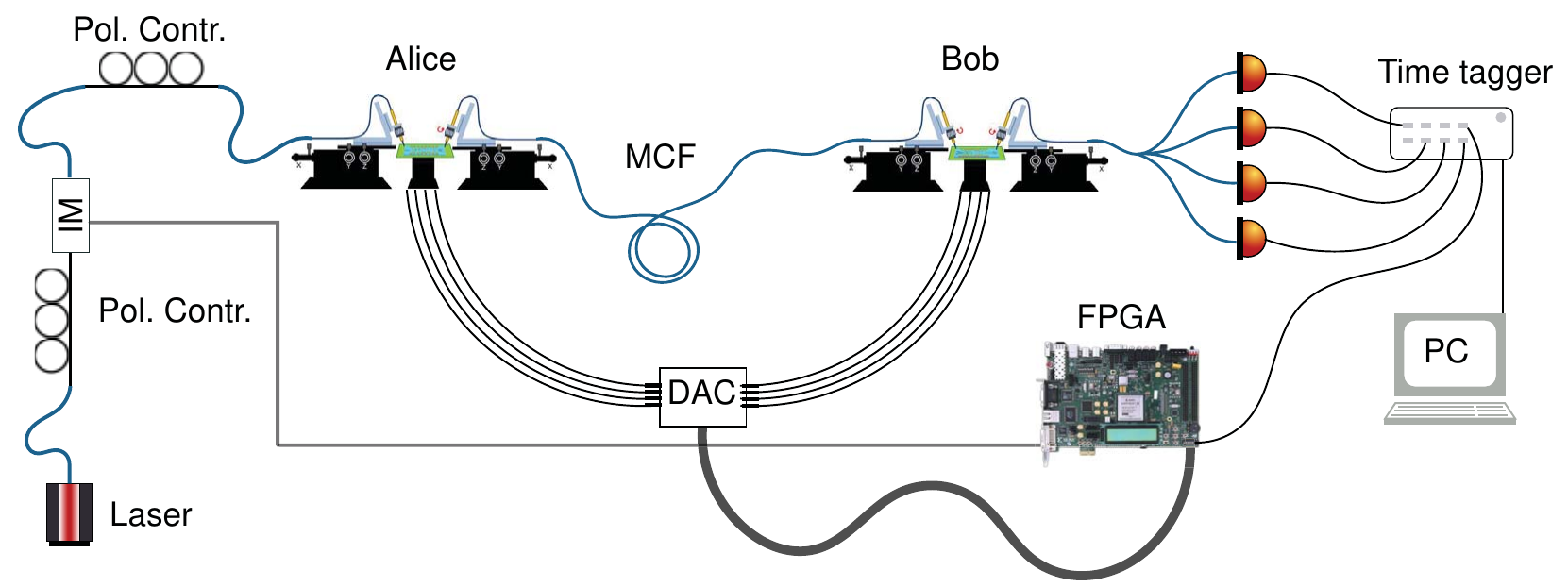}
\caption{\textit{Experimental Setup}. A $1540$ nm CW fiber laser was used for an HD-QKD experiment.  An intensity modulator, controlled by an FPGA was used to generate a $5$ kHz train of pulses. In order to create the quantum states, the light was coupled in Alice chip by using analog electrical signals, generated by the FPGA and DAC. In this way it was possible to prepare WCPs based on spatial division multiplexer. A PRBS sequence ($8$ bit) was used to change randomly the Alice's states and bases encoding. After $3$ meters of MCF we coupled the light into Bob. The receiver can measure the quantum states in different bases depending on the the signals coming from the DAC chips. In this case we used an independent PRBS (7 bit) to choose between the different bases. Finally, after coupling the $4$ different outputs into $4$ single mode fibers we detected photons using free-running ID230 and ID220 InGaAs SPDs.}
\label{fig:setup}
\end{figure}

We also report in Fig. S~\ref{fig:pulsewidth} the accumulated counts acquired with a time-tagger for 45 seconds, indicating a pulse  width of 10 ns. The counts outside the pulse width indicates the limited modulation extinction ratio.

\begin{figure}[!t]
\begin{center}
\includegraphics[width=0.6\textwidth]{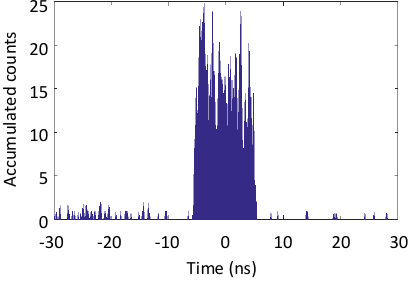}
\end{center}
\caption{\textit{Accumulated counts of WCPs.} Pulse width of accumulated counts for $45$ seconds using $10\%$ detector efficiency and $20 \mu$s of dead time. In this configuration, the jitter introduced by the detectors is around $400$ ps. Average pulse width of $10$ ns represents a strong advantage in a slow quantum system (order of kHz), where the noise of the channel and detectors compromise your communication.}
\label{fig:pulsewidth}
\end{figure}

\begin{figure}[!t]
\begin{center}
\includegraphics[width=16cm]{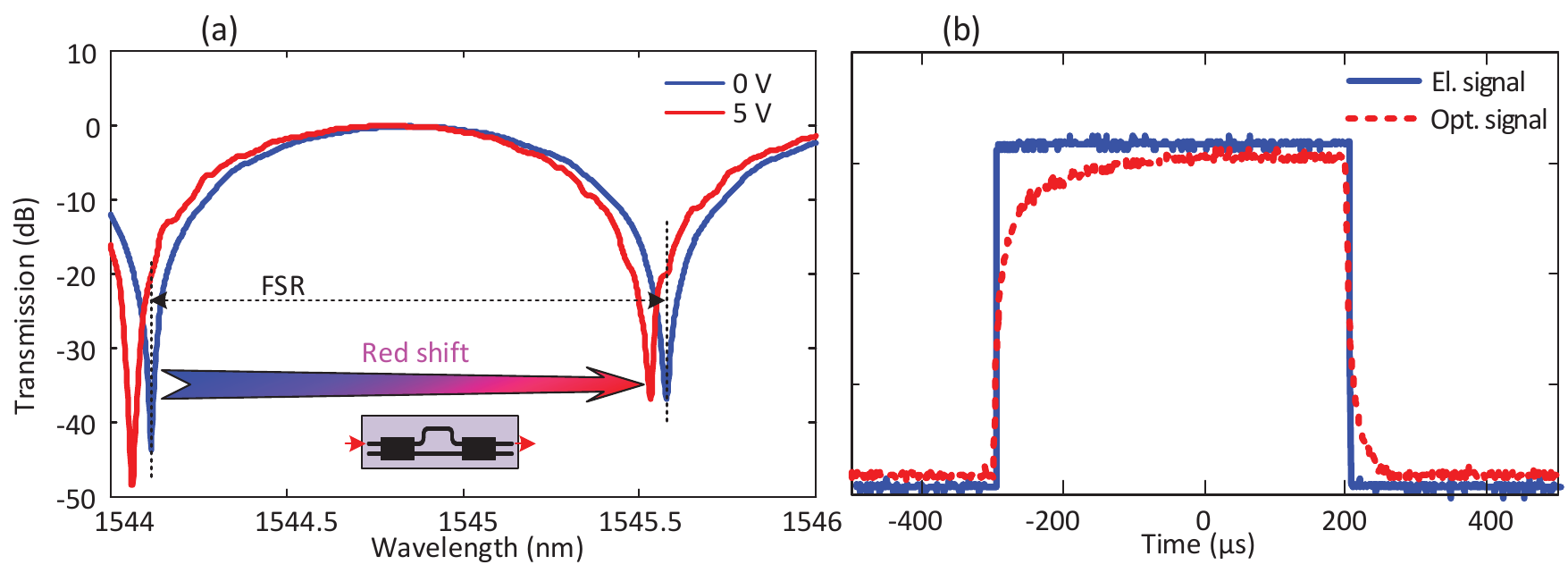}
\end{center}
\caption{\textit{Characterization of heater response by a tunable MZI}. (a) Transmission of MZI for different applied voltage. (b) Switching speed characterization by applying electrical square form on the heaters, and measuring the switched light power, indicating a switching time of 66 $\mu$s and 27 $\mu$s for rise ($10\%$ to $90\%$) and fall time ($90\%$ to $10\%$), respectively.}
\label{fig:Heater}
\end{figure}

\subsection*{Characterization of switching time}
The thermal switching time was characterized by fabricating an asymmetrical MZI with the same heater design that was used for Alice and Bob, introduced in one of the two arms. Figure S.~\ref{fig:Heater}(a) shows normalized MZI transmission spectra for different driving voltage. A $5$V driving voltage results in an almost one FSR shift. The maximum driving voltage from the FPGA is $4.1$V, thus a $\pi$ phase shift is guaranteed. The dynamic switching response is exhibited in Figure S3(b), where a square electrical waveform at $2$kHz was applied on the MZI, the output switched optical signal was detected. A switching time of 66 $\mu$s and 27 $\mu$s for rise ($10\%$ to $90\%$) and fall time ($90\%$ to $10\%$), respectively was obtained. The time of being stabilized ($0\%$ to $99\%$) is $~180\mu$s, which restricts the repetition rate in the experiment.

\subsection*{Mutual information}
In the most common BB84 protocol the threshold of the QBER is fixed to $11\%$ considering one way information reconciliation. Values higher than the threshold interrupt the key generation process. However, in case of qudit encoding, this value can be increased. In fact, as reported in Figure S.~\ref{img:IabIae}, the intersection point between the solid and dashed curves represents the disturbance limit on the QKD d-level system.
These limits are derived considering the case of $2$ MUBs and under the hypothesis of individual attacks. The limit value is slightly increased in the case of a complete set of MUBs, defined as $N+1$ MUBs bases, where $N$ is the Hilbert space dimension. In Table S.1 we reported the various threshold values for different $N$ dimension and attacks. 

\begin{table}[!h]
\centering
\label{table:limits}
\caption{Disturbance limit for different Eve's strategy and various Hilbert space dimension. Note that $N=6$ is missing from the list because $6$ is not a prime number (or a multiple of a prime number) and $N+1$ bases can not be defined. 
}
\begin{tabular}{ l c c  r }
  N & $D_2^{ind}$ & $D_{N+1}^{ind}$ & $D_{coh}$\\
  \addlinespace[0.5ex]
  \hline			
  \addlinespace[0.5ex]
  2 & 14.64 & 15.64 & 11.00  \\
  3 & 21.13 & 22.67 & 15.95\\
  4 & 25.00 & 26.66 & 18.93 \\
  5 & 27.64 & 29.23 & 20.99\\
  8 & 32.32 & 33.44 & 24.70\\
  \hline  
\end{tabular}
\end{table}

\begin{figure}[!b]
\centering
\includegraphics[width=0.6\textwidth]{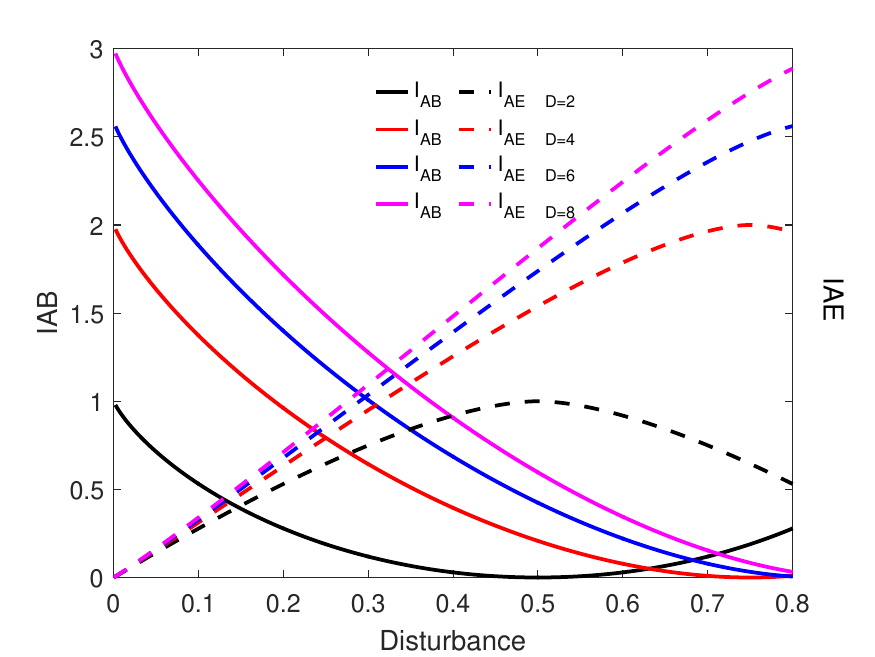}
\caption{\textit{Mutual information for different Hilbert space.} Under the hypothesis of individual attacks and $2$ MUBs we reported the theoretical bound of mutual information between Alice and Bob/Eve for different Hilbert space dimension. The intersection between solid and dashed curves represents the disturbance threshold after which no key can be extracted.}
\label{img:IabIae}
\end{figure}


\begin{figure}[!ht]
\includegraphics[width=1\textwidth]{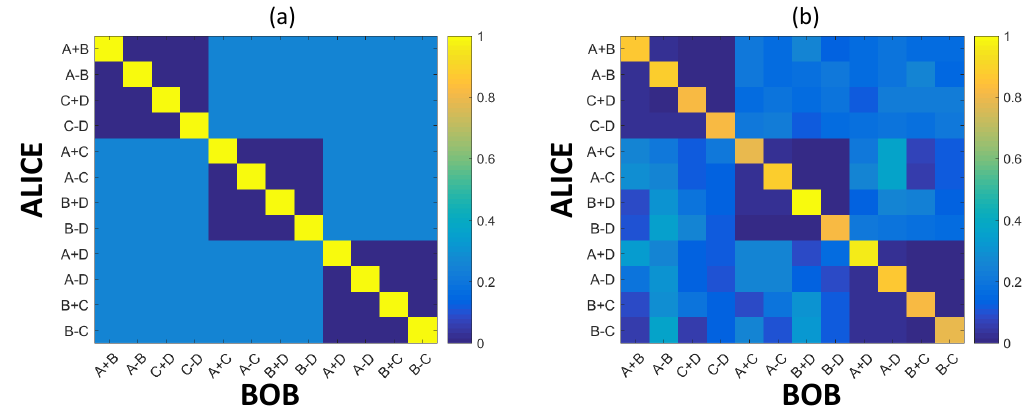}
\caption{\textit {Two dimensional view of the three MUBs}. Theoretical on the left and experimental on the right representation of the $3$ MUBs. The matrix was acquired with a $1540$ nm, $10$ kHz repetition rate, $10$ns wide laser. We used low mean photon per pulse $\mu$, around $0.2$ and each pixel was obtained with several measurement of $30$ s slot with a total number of pulses of $34$ million.  }
\label{fig:simulation}
\end{figure}

\end{document}